\documentclass[superscriptaddress,showpacs,twocolumn,pre]{revtex4}
\usepackage{graphicx}
\usepackage{bm}
\usepackage{amsmath}
\newcommand{\bkt}[1]{\left\langle#1\right\rangle}

\begin{document}

\title{Acceleration Statistics as Measures of Statistical 
Persistence of Streamlines in Isotropic Turbulence}

\author{S. Goto}
\altaffiliation[Permanent address: ]{Dept. Mechanical Engineering, Kyoto 
University, Kyoto 606-8501 Japan.}
\affiliation{
Turbulence and Mixing Group,
Department of Aeronautics, Imperial College, London, 
SW7 2AZ, UK}

\author{D.R. Osborne}
\affiliation{
Turbulence and Mixing Group,
Department of Aeronautics, Imperial College, London, 
SW7 2AZ, UK}
\affiliation{
Space and Atmospheric Physics, 
Department of Physics, Imperial College, London, 
SW7 2AZ, UK}

\author{J.C. Vassilicos}
\affiliation{
Turbulence and Mixing Group,
Department of Aeronautics, Imperial College, London, 
SW7 2AZ, UK}

\author{J.D. Haigh}
\affiliation{
Space and Atmospheric Physics, 
Department of Physics, Imperial College, London, 
SW7 2AZ, UK}

\date{\today}

\begin{abstract}
We introduce the velocity $\bm{V}_s$ of stagnation points as a means
to characterise and measure statistical persistence of streamlines. Using 
theoretical arguments, Direct Numerical Simulations (DNS)
and Kinematic Simulations (KS) of three-dimensional isotropic
turbulence for different ratios of inner to outer length scales
$L/\eta$ of the self-similar range, we show that a frame exists where
the average $\bkt{\bm{V}_s}=\bm{0}$, that the r.m.s.~values of
acceleration, turbulent fluid velocity and $\bm{V}_s$ are related by
$La'/u'^{2} \sim (V_{s}'/u') (L/\eta)^{2/3+q}$ and that $V_{s}'/u'
\sim (L/\eta)^{q}$ with $q=-1/3$ in Kolmogorov turbulence, $q=-1/6$ in
current DNS and $q=0$ in our KS. The statistical persistence
hypothesis is closely related to the Tennekes sweeping hypothesis.
\end{abstract}


\maketitle

Accelerations are central to fluid flow.  Acceleration 
statistics are central to turbulence dynamics, turbulent mixing and 
transport, and also important where the turbulence controls droplet growth, 
chemical reactions, combustion and other processes 
(e.g.~\cite{Shaw, Pope}). A key statistic is the acceleration 
variance. Its value and scaling with Reynolds number are essential to 
stochastic Lagrangian models and to Lagrangian probability density 
function models of turbulent diffusion 
if these models are to incorporate finite Reynolds number effects 
\cite{Pope}.
In this letter we argue that the Reynolds number scaling of the
turbulence acceleration statistics, in particular the acceleration
variance, are also important as measures of the statistical
persistence of streamlines.

The persistence of streamlines is central to turbulent
pair diffusion \cite{FHMP92, FV98, DV, GV}. It may also
be important for one-particle two-time Lagrangian turbulent statistics
\cite{KV04}. If streamlines are persistent
enough in time, fluid element trajectories approximately follow
them for significantly long times. Hence, initially
close fluid element pairs typically separate when they encounter a
region of highly curved diverging streamlines around a straining
stagnation point. However, streamlines and their persistence are not
Galilean invariant. This issue motivated \cite{GV} to suggest that the
dependence of turbulent diffusion on streamline structure fully
emerges only in the frame of reference where streamline persistence is
maximised in some statistical sense. The assumption that such a frame
exists, and that the statistical persistence of the streamline
topology in this frame is long enough to leave its defining imprint on
turbulent diffusion, they termed 'statistical persistence
hypothesis'. This is a powerful hypothesis as it leads to predictions
supported by DNS and KS such as $\gamma = 2d/D_{s}$ (see \cite{DV,
GV}) which relate the Richardson pair-diffusion exponent $\gamma$ (a
Lagrangian quantity) in $d$-dimensional turbulence to the fractal
dimension $D_s$ of the spatial distribution of stagnation points (an
Eulerian quantity) in the privileged frame of the statistical
persistence hypothesis. For isotropic turbulence, \cite{GV} argued
that this frame is the one where the mean fluid velocity vanishes. The
purpose of this letter is to mathematically formulate the statistical
persistence of streamlines as well as the statistical persistence
hypothesis (statistical persistence of streamlines maximised in a
well-chosen frame) and to argue that, in isotropic turbulence, they
are both increasingly valid for increasing Reynolds number.  

It is the persistence of highly curved diverging streamlines around
straining stagnation points which is important for turbulent
diffusion, and it is therefore critical to consider the speed with
which stagnation points move in space. Given an arbitrary frame of
reference, the fluid velocity $\bm{u}$ at a stagnation point 
$\bm{s}(t)$ at time $t$ vanishes, i.e.~$\bm{u}(\bm{s},t)=\bm{0}$, and
remains so for as long as this stagnation point exists. Hence, during
the stagnation point's life-time, $\bm{0}={d\over dt}\bm{u}(\bm{s}, t)
={\partial\bm{u}\over\partial t}+ \bm{V}_s\cdot\nabla
\bm{u}$ at position $\bm{s}$ and time $t$, and $\bm{V}_s \equiv {d\bm{s}\over
dt}$ is the stagnation point velocity. 
The fluid acceleration is defined as $\bm{a} \equiv
{\partial\bm{u}\over\partial t} + \bm{u}\cdot\nabla\bm{u}$
at all positions $\bm{x}$ and times $t$. Setting $\bm{x} =\bm{s}$,
we obtain
\begin{equation}
 \bm{a} = - \bm{V}_s\cdot\nabla\bm{u}
 \label{aVs}
\end{equation}
at any time $t$ and any stagnation point $\bm{s}(t)$ of the
flow. Using Cramer's rule, and assuming that 
$det({\partial\bm{u}\over\partial x}, 
     {\partial\bm{u}\over\partial y}, 
     {\partial\bm{u}\over\partial z})\ne0$, this equality can be inverted:
\begin{align}
 \bm{V}_s 
 = 
 -
 \frac{
 \left[
  det(\bm{a},
      {\partial\bm{u}\over\partial y},
      {\partial\bm{u}\over\partial z}), 
  det({\partial\bm{u}\over\partial x}, 
      \bm{a}, 
      {\partial\bm{u}\over\partial z}), 
  det({\partial\bm{u}\over\partial x}, 
      {\partial\bm{u}\over\partial y}, 
      \bm{a})
  \right]
 }
 {
  det({\partial\bm{u}\over\partial x}, 
      {\partial\bm{u}\over\partial y}, 
      {\partial\bm{u}\over\partial z})
 }.
\label{Vs=}
\end{align}
Note that relations (\ref{aVs}) and (\ref{Vs=}) hold in any frame of
reference. What changes with frame is the number and positions of
stagnation points (where these relations hold).

These relations are kinematic and are the fundamental link we use here
between accelerations and the persistence of stagnation points
measured by $\bm{V}_s$.  We formulate the concept of statistical
persistence of streamlines on the basis of (\ref{aVs}). Assuming, as
is reasonable in fully developed isotropic turbulence \cite{Pearson}, that the
kinetic energy dissipation rate per unit mass, $\epsilon$, scales as
$u'^{3}/L$ (where $u'$ is the turbulence velocity fluctuations r.m.s.,
and $L$ is an outer length-scale of the turbulence) and that the
velocity gradients in (\ref{aVs}) scale with the inner length-scale
$\eta$ and the small-scale velocity $u_{\eta}\sim (\epsilon
\eta)^{1/3}$, then the acceleration r.m.s.~$a'$ is related to the
r.m.s.~$V_{s}'$ of $\bm{V}_s$ by
\begin{equation}
V_{s}'\sim a' \tau_{\eta}
\label{Vtau}
\end{equation}
where $\tau_{\eta}\sim \eta/u_{\eta}$, or equivalently
\begin{equation}
 La'/u'^{2} \sim (V_{s}'/u')(L/\eta)^{2/3}.  \label{a'}
\end{equation}
Strictly, $a'$ is the acceleration r.m.s. over all stagnation points, 
but it is also equal to the acceleration r.m.s. over the entire field 
because Galilean transformations leave $a'$ unchanged even though they 
cause the r.m.s. statistics to be calculated over different ensembles 
of points.  Our DNS and KS calculations of $a'$ and $V_s'$ support this 
view.  The Kolmogorov scaling of $a'$ (where $a'$ is determined by 
$\epsilon$ and $\eta$) is $La'/u'^{2} \sim (L/\eta)^{1/3}$ and has been
corroborated (using $\eta \sim (\nu^{3}/\epsilon)^{1/4}$ where $\nu$
is the kinematic viscosity) by laboratory measurements of acceleration
statistics in isotropic turbulence \cite{Boden} in the range
$900<Re_{\lambda}<2000$, where $Re_{\lambda}$ is the
Taylor-length-based Reynolds number. This scaling implies $V_{s}'/u'
\sim (L/\eta)^{-1/3}$ which means that the movement of stagnation points 
(characterised by
$V_{s}'$) is, statistically, increasingly slower than that of fluid elements
(characterised by $u'$) as the Reynolds number increases. In this
statistical sense, streamline curvature around stagnation points
becomes increasingly persistent and fluid element trajectories can
follow the curvature of these streamlines for an increasingly
significant time.  This interpretation assumes, as we confirm in our
DNS and KS studies of isotropic turbulence which we report below, that
in the frame where the mean flow vanishes, the average of $\bm{V}_s$
over the entire flow is $\bm{0}$, i.e.~$\bkt{\bm{V}_s}=\bm{0}$. In
other inertial frames, $\bkt{\bm{V}_s}$ is not zero but proportional
to the velocity $\bm{U}$ of the frame relative to the one where the
mean flow vanishes, thus reducing the persistence of streamlines in
such other frames.  (Stagnation points in these frames $F$ correspond
to points with fluid velocity $\bm{U}$ in the frame $F_0$ where
$\bkt{\bm{V}_s}=\bm{0}$.  The average acceleration over these points
in frame $F_0$ is $\bkt{\bm{a}\vert\bm{U}}$, i.e.~the average of
$\bm{a}$ conditional on points where the fluid velocity is $\bm{U}$,
and is equal to the average acceleration over the corresponding
stagnation points in frame $F$ which follows from (\ref{aVs}) and is
given by $-\bkt{\bm{V}_s\cdot\nabla\bm{u}}$.  The DNS study which we
report below indicates that $\bkt{\bm{a}\vert\bm{U}}\propto \bm{U}$
for all Reynolds numbers tested, thus suggesting
$\bkt{\bm{V}_s}\propto\bm{U}$, a proportionality relation confirmed by
our DNS with a negative proportionality coefficient.)  In conclusion,
a frame exists where the statistical persistence of
streamlines is maximised and where the statistical persistence
hypothesis is valid provided that $V_{s}'$ is much smaller than $u'$;
this frame is the one where the average of $\bm{V}_s$ is zero, hence
the one where the mean flow vanishes as argued by \cite{GV}.

DNS studies of three-dimensional isotropic turbulence support the
non-Kolmogorov scaling $La'/u'^{2} \sim (L/\eta)^{1/2}$ in the range
$40<Re_{\lambda}<230$ (see \cite{Vedula} and note that $\epsilon \sim
u'^{3}/L$ is approximately valid in such simulations with such
Reynolds numbers \cite{Pearson} where, also, $\eta \sim
(\nu^{3}/\epsilon)^{1/4}$) and therefore imply $V_{s}'/u' \sim
(L/\eta)^{-1/6}$ in that range. This also means that stagnation points
become increasingly persistent as Reynolds number increases, but at a
rate slower than that of Kolmogorov turbulence. Defining an exponent, 
$q$, by
\begin{equation}
La'/u'^{2} \sim (L/\eta)^{2/3 + q} \sim Re_{\lambda}^{1+3q/2}
\label{aq}
\end{equation}
(where $Re_\lambda\sim (L/\eta)^{\frac{2}{3}}$ has been used), it follows 
from (\ref{a'}) that (\ref{aq}) is equivalent to 
\begin{equation}
V_{s}'/u' \sim (L/\eta)^{q} \sim Re_{\lambda}^{3q/2}; 
\label{Vq}
\end{equation}
$q>0$ corresponds to absence and $q<0$ corresponds to presence of
statistical persistence of streamlines (in the sense that frames exist
where streamlines are statistically fairly persistent and a
privileged frame also exists --- the one where $\bkt{\bm{V}_s}=\bm{0}$ ---
where this statistical persistence is maximised and the statistical
persistence hypothesis holds). In the remainder of this letter we use
DNS and KS to confirm and further explore the behaviour of $\bm{V}_s$,
the Reynolds number scalings of $La'/u'^{2}$ and $V_{s}'/u'$ and the
relation between these scalings via (\ref{a'}). An advantage of
formula (\ref{Vs=}) is that it can be used to calculate $\bm{V}_s$
from Eulerian snapshots of DNS and KS velocity fields without having
to track the motion of stagnation points.

\begin{figure*}
\includegraphics[height=4.6cm]{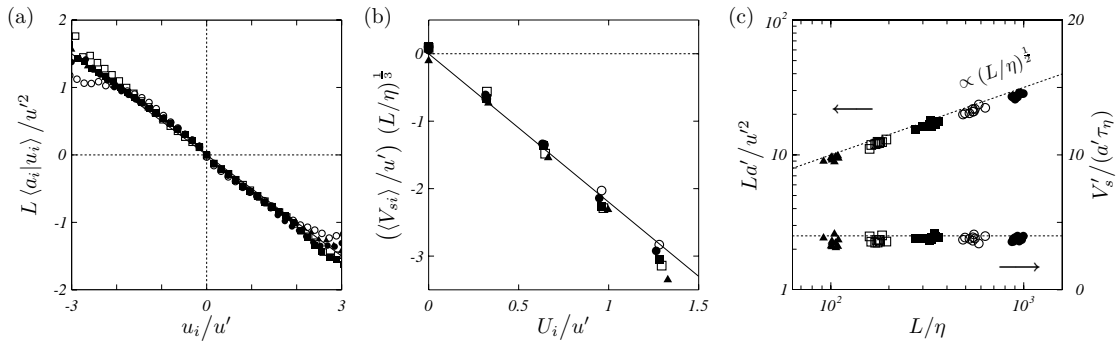}
\caption{\label{fig:dns} Statistics of acceleration and stagnation
point velocity. DNS. Solid triangles, $Re_\lambda=57$; 
open squares, $83$;
solid squares, $120$;
open circles, $180$;
solid circles, $250$.
(a) Conditional average of acceleration in the frame $F_0$.
(b) Average stagnation point velocity in the moving frame with $\bm U$ 
relative to $F_0$.
(c) R.m.s.~acceleration and r.m.s.~stagnation velocity; here we plot
results of ten different snapshots from each simulation. $L/\eta$ 
fluctuates in time which explains the small scatter around its 
average.  
}
\end{figure*}

We use DNS data of non-decaying homogeneous isotropic
three-dimensional incompressible turbulence generated by a standard
spectral code (the magnitudes of a few low wavenumber modes are kept constant) 
with grid resolution of about $2\eta$ (except for the
highest Reynolds number where it is about $3\eta$) and compute
stagnation points and their velocities $\bm{V}_s$ in instantaneous
velocity fields for $Re_{\lambda}$ ranging from $57$ to $250$ (see
\cite{GK03} for the DNS numerical scheme and parameters). We use the
Newton-Raphson method to find all stagnation points. This is an
iterative method and requires starting points, which have been taken
over the DNS field's entire domain at points separated by the
numerical grid width. For every Reynolds number, we calculate
$\bkt{\bm{V}_s}$ in various frames $F$, $\bkt{\bm{a}\vert\bm{u}}$ in
frame $F_0$ and also $a'$, $V_{s}'$, the integral length-scale $L$ and $\eta$ 
(we calculate $-\bm{\nabla}p+\nu\nabla^2\bm{u}$ to obtain $\bm{a}$.  If we 
wrongly approximate $\bm{a}=-\bm{\nabla}p$, then we find 
$\bkt{\bm{a}\vert\bm{u}}=0$). 
Our results show for all our Reynolds numbers, that
$\bkt{\bm{a}\vert\bm{u}}\approx-{u'\over 2L} (\bm{u} - \bkt{\bm{u}})$
(see Fig.~\ref{fig:dns}(a); note that we observe components
$\bkt{a_{i}\vert u_{j}}$ to be 0 only if $i \ne j$) and that
$\bkt{\bm{V}_s}=\bm{0}$ only when $\bm{U}=\bm{0}$ (see
Fig.~\ref{fig:dns}(b)). We also plot $La'/u'^{2}$ and
$V_{s}'/(a'\tau_{\eta})$ as functions of $L/\eta$ (see
Fig.~\ref{fig:dns}(c)). In agreement with previous DNS studies (see
\cite{Vedula} and references therein) we obtain $La'/u'^{2} \sim
(L/\eta)^{1/2}$ (i.e.~$q=-1/6$) and confirm 
(\ref{Vtau}) over our DNS range of Reynolds numbers. By the way, from
$\bkt{\bm{a}\vert\bm{U}}=-\bkt{\bm{V}_s\cdot\nabla\bm{u}}$ and
$\bkt{\bm{a}\vert\bm{U}}\approx-{u'\over 2L}\bm{U}$ one might expect
$\bkt{\bm{V}_s}\propto\bm{U}$ with a constant of proportionality of
decreasing absolute value for increasing $Re_{\lambda}$, as we indeed
observe.  What is unexpected (but is left for future
investigation) is the rate of this decrease;
Fig.~\ref{fig:dns}(b) suggests 
$\langle\bm{V}_s\rangle = -b (L/\eta)^{-1/3} \bm{U}$
where $b$ is a positive dimensionless universal constant. At any rate,
our DNS suggest that $\bkt{\bm{V}_s}/u'\to\bm{0}$ and $V_{s}'/u' \to
0$ as $Re_{\lambda}\to \infty$. From (\ref{aVs}), this implies
that as $Re_{\lambda}\to \infty$, stagnation points tend to
become non-moving zero-acceleration points.


To gain further insight into $\bkt{\bm{V}_s}$ and the validity of
relations (\ref{a'}), (\ref{aq}) and (\ref{Vq}) and how they depend on
specific properties of the underlying flow, we now study synthetic
velocity fields, namely KS, where the spectrum and the time-dependence
can be modified at will \cite{FHMP92, FV98, DV, OVH}. Such a study
cannot be carried out with DNS where the spectrum and time-dependence
of the flow are determined by the Navier-Stokes dynamics and cannot be
tampered with.  An additional advantage of KS is that the Lagrangian
statistics it produces compare well with various DNS and laboratory
results (see \cite{OVH} and references therein).  Finally, because of
the dramatic decimation in number of modes, it is possible with KS to
explore scalings with $L/\eta$ up to extremely large $L/\eta$ values
(here $10^6$) which are out of reach of current DNS.

In our KS we use three-dimensional turbulent-like velocity fields of the form 
(see \cite{OVH} for fuller details)
\begin{eqnarray}
 \bm{u}
 =
 \sum_{n=1}^{N_k}
   \bm{A}_n 
   \cos(\bm{k}_n\cdot\bm{x} + \omega_n t) 
   + \nonumber 
   \bm{B}_n 
   \sin(\bm{k}_n\cdot\bm{x} + \omega_n t)  
\label{KS}
\end{eqnarray}
where $N_k$ is the number of modes, 
the directions and orientations of ${\bm A}_n$ and ${\bm B}_n$ are
chosen randomly 
and uncorrelated with the directions and orientations of all other
wave modes but perpendicular to $\bm{k}_n$.  The distribution of
wavenumbers is geometric, specifically $ k_n \equiv  \vert \bm{k}_n \vert 
= k_1 1.07^{n-1}$.  The
velocity field is incompressible by construction, and also
statistically stationary, homogeneous and isotropic as shown by
\cite{FHMP92, FV98}. The amplitudes of the vectors ${\bm A}_n$ and
${\bm B}_n$ are determined from the energy spectrum $E(k_{n})$ 
prescribed to be of the form
\begin{equation}
   E(k) = \frac{3(p-1)\:u'^{2}}{2\,(L/2\pi)^{p-1}}\:k^{-p} 
\end{equation}
in the range $ 2\pi/L = k_1 \le k \le k_{N_k} = 2\pi/\eta $, and $
E(k)=0 $ otherwise.
The ratio $L/\eta$ is increased by increasing $N_{k}$. 
Following \cite{FV98, OVH}, we
set $\omega_n = \lambda u' k_n$, for different values of the
dimensionless parameter $\lambda$.

From (\ref{aVs}) we can derive a generalised form of (\ref{a'}) within
the framework of KS by assuming that the velocity gradients in
(\ref{aVs}) scale with the inner length-scale $\eta$ and the
small-scale velocity $u_{\eta}\sim u'(\eta/L)^{{p-1\over 2}}$. This leads to
$La'/u'^{2} \sim (V_{s}'/u')(L/\eta)^{{3-p\over 2}}$ which generalises
(\ref{a'}) and to the statement that $La'/u'^{2} \sim
(L/\eta)^{{3-p\over 2}+q}$ is equivalent to $V_{s}'/u' \sim
(L/\eta)^{q}$ which generalise (\ref{aq}) and (\ref{Vq}).

\begin{figure*}
\includegraphics[height=4.6cm]{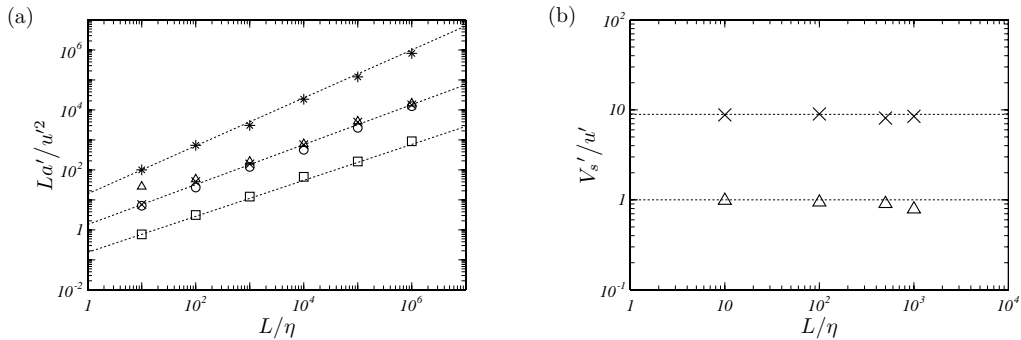}
\caption{\label{fig:ksp} Statistics of acceleration and stagnation
point velocity. KS. (a) R.m.s. acceleration scaling with ${L}/{\eta}$.  
For $p={5}/{3}$; crosses, $\lambda=0$; circles, $0.5$; triangles, $5.0$
For $\lambda=0.5$; asterisks, $p=1.4$; squares, $1.8$.  
(Data sets for these $p$ are shifted with respect to the vertical axis
by factor $10$ for $p=1.4$ and by $0.1$ for $1.8$.)
 The slopes of the lines are ${(3-p)}/{2}$.
(b) Variation of r.m.s stagnation point velocity with ${L}/{\eta}$.  
Triangles, $\lambda=0.5$; crosses, $\lambda=5.0$.
}
\vspace{-3mm}
\end{figure*}

KS runs with $p=5/3$, values of $L/\eta$ ranging between
$10$ and $10^3$, and $\lambda = 0, 0.5, 5$, all lead to
$\bkt{\bm{a}\vert\bm{u}}=\bm{0}$ and to $\bkt{\bm{V}_s}=\bm{0}$ in
frame $F_0$.  In frames $F$, $\langle {\bm{V}_s}\rangle=-{\bm{U}}$.  In 
KS, ${\bm{V}_s}$ is uncorrelated with ${\nabla}\bm{u}$ so that 
$\langle{\bm{V}_s}\cdot{\nabla}{\bm{u}}\rangle
=\langle{\bm{V}_s}\rangle\cdot\langle{\nabla}{\bm{u}}\rangle=0$ 
in agreement with
$\bm{0}=\bkt{\bm{a}\vert
\bm{U}}=-\bkt{\bm{V}_s\cdot\bm{\nabla}\bm{u}}$. Runs with $p=1.4, 5/3,
1.8$, values of $L/\eta$ 
ranging between $10$ and $10^6$, and $\lambda = 0, 0.5, 5$ also lead to
$La'/u'^{2} \sim (L/\eta)^{{3-p\over 2}}$ (see Fig.~\ref{fig:ksp}(a)) and
therefore $q=0$, and to $V_{s}'/u' = c$ where $c$ is a dimensionless
constant independent of $L/\eta$, which confirms that $q=0$
(see Fig.~\ref{fig:ksp}(b)). (In all our KS cases, $5\times10^{6}$ starting
points for the Newton-Raphson method are chosen over the same volume
$L_s^{3}$ where $L_s$ is the $L$ corresponding to the largest $L/\eta$
tried, keeping $\eta$ constant.)

The result $\bkt{\bm{a}\vert\bm{u}}=\bm{0}$ reflects the
lack of dynamics and related lack of correlations between
Fourier modes in KS. Turbulence dynamics seem to generate restoring
accelerations which are anti-correlated with velocities (see
Fig.~\ref{fig:dns}(a)).

The constant $c$ turns out to be an increasing function of 
$\lambda$ (from our simulations,
$c\propto \lambda$) and, as expected, $c=0$ for $\lambda
=0$. Hence, in our KS isotropic turbulence where $q=0$, the
statistical persistence of streamlines is measured by $c$ and is a
direct reflection of the unsteadiness parameter $\lambda$. In spite of
$q$ being different from the Kolmogorov value $-1/3$, Richardson
exponents $\gamma = 2d/D_{s}$ (in particular $\gamma =3$ for $p=5/3$)
are observed in KS but only for small values of $\lambda$ \cite{FV98,
DV} thus confirming the view that these exponents require some
statistical persistence of streamlines to be realised \cite{GV}.

Finally, it is worth recalling the Tennekes sweeping hypothesis
\cite{tennekes} which states that the
dissipative eddies are swept past an Eulerian observer in a time much
shorter than the time scale characterizing their own
dynamics. The statistical persistence hypothesis, the validity of
which we confirm in this letter, states that there exists a frame
where $\bkt{\bm{V}_s}=\bm{0}$ and $V_{s}'\ll u'$; from (\ref{Vtau}),
it therefore follows that $a' \ll u'/\tau_{\eta}$. In this sense, the
accelerations are small, which is a way to restate the Tennekes
sweeping hypothesis. Indeed, the time needed for
dissipative eddies to sweep past an Eulerian observer is $\eta/u'$
which is therefore much smaller than $u_{\eta}/a'$, the time which
characterizes the dynamics of these eddies. This is the Tennekes
sweeping hypothesis derived from the statistical persistence of
streamlines. Alternatively, stagnation points mark regions of the flow
where there is no sweeping. According to \cite{tennekes}, statistics
which are taken so as to remove the sweeping effect depend only on the
small-scale dynamics, and this must therefore be the case of
$V_{s}'$. In Kolmogorov turbulence, the scaling of these small-scale
dynamics is determined by $\epsilon$ and $\nu$ which therefore implies
$V_{s}'\sim (\epsilon \nu)^{1/4}$ in agreement with $q=-1/3$ and the
statistical persistence hypothesis. In the present KS, however,
sweeping of smaller-scale turbulence by larger-scale eddies is absent,
and stagnation points correspond to regions where zeros of Fourier
modes congregate. These Fourier modes move together with velocity
$\omega (k)/k$. Hence, stagnation points typically move with that same
speed and $V_{s}' \sim \omega (k)/k \sim \lambda u'$ in agreement with
our KS results $q=0$ and $c\propto \lambda$. 

The statistical persistence of streamlines seems to be a reformulation
and generalisation of the Tennekes sweeping hypothesis in terms of
streamline topology and its persistence.  Also, in conjunction with the kinematic 
relation (\ref{aVs}), Kolmogorov 
dimensional analysis implies the existence of a coherent flow structure, 
namely the persistent multiple-scale stagnation point structure of the 
turbulence.  Elsewhere \cite{iutam} we argue that the mean life-time of 
stagnation points is of the order of the integral time-scale of the flow.

\vspace{-8mm}

\begin{acknowledgments}
We acknowledge support from the Japanese Ministry of
Education, Culture, Sports, Science and Technology; NERC; and 
the Royal Society of London.
\end{acknowledgments}

\end{document}